\documentclass[aps,preprint,epsfig,rotate]{revtex4}
\begin{document}
\title{Annihilation, bound state properties and photodetachment of the positronium negatively charged ion.}

\author{Alexei M. Frolov}
 \email[E--mail address: ]{afrolov@uwo.ca}

\affiliation{Department of Applied Mathematics \\
 University of Western Ontario, London, Ontario N6H 5B7, Canada}

\date{\today}

\begin{abstract}

Bound state properties of the negatively charged Ps$^{-}$ ion (or $e^{-} e^{+} e^{-}$) are discussed. The expectation values of operators which correspond to these 
properties have been determined with the use of the highly accurate wave functions constructed for this ion. Our best variational energy obtained for the Ps$^{-}$ 
ion is $E$ = -0.2620050 7023298 0107770 40051 $a.u.$ Annihilation of the electron-positron pair(s) in the negatively charged Ps$^{-}$ ion (or $e^{-} e^{+} e^{-}$) 
is considered in detail. By using accurate values for a number annihilation rates $\Gamma_{n \gamma}$, where $n$ = 1, 2, 3, 4 and 5, we evaluated the half-life $\tau_a$ 
of the Ps$^{-}$ ion against positron annihilation ($\tau_a = \frac{1}{\Gamma} \approx 4.793584140 \cdot 10^{-10}$ $sec$). Photodetachment of the Ps$^{-}$ ion is considered 
in the long-range, asymptotic approximation. The overall accuracy of our photodetachment cross-section is very good for such a simple approximation. \\

\noindent 
PACS number(s): 36.10.-k and 36.10.Dr

\noindent 
First version 30.12.2014, Preprint-2014-15/1 (this is 4th version) [at.phys.], 16 pages.

\end{abstract}

\maketitle


\section{Introduction}

The main goal of this short communication is to perform computational and theoretical analysis of the annihilation of electron-positron pair(s) in the negatively charged Ps$^{-}$ 
ion. Another aim is to evaluate the photodetachment cross-section of the Ps$^{-}$ ion in the long-range, asymptotic approximation. Our analysis is based on some recent results 
of highly accurate computations performed for the ground (bound) $1^1S-$state in the Ps$^{-}$ ion, which is also designated as the $e^{-} e^{+} e^{-}$ ion, or $e^{+} e^{-}_{2}$ ion. 
Stability of this three-body system was predicted by Ruark \cite{Ruar}. First variational calculations of the ground state in the Ps$^{-}$ ion were performed by Hylleraas in 1947 
\cite{Hyll}. This ion is of interest in various branches of physics, including solid state physics \cite{Gidley}, astrophysics \cite{Dra}, \cite{BD2}, \cite{BD3}, physics of 
high-temperature plasmas, etc. Note that the Ps$^{-}$ ion has been created experimentally by Mills almost forty years ago \cite{Mills}. Review of the most recent experiments 
performed for the $Ps^{-}$ ion can be found in \cite{Nag} and \cite{Klaus} which also contain a large number of useful references. 

As is well known from the general theory of bound states in the Coulomb three-body systems with unit charges \cite{Fro92} this ion has only one stable state, which is the ground 
$1^1S-$state, or $1^1S(L = 0)$-state. In general, to describe the bound $1^1S-$state in this ion we can restrict ourselves to the non-relativistic $LS-$approximation, since all
lowest-order relativistic and QED corrections are very small for this ion. The non-relativistic approximation means that the wave function $\Psi$ can be determined as the solution 
of the non-relativistic Schr\"{o}dinger equation $H \Psi = E \Psi$, where $E < 0$, for three-particle quasi-atomic (or Coulomb) systems. The non-relativistic Hamiltonian $H$ of the 
Ps$^{-}$ ion takes the form
\begin{eqnarray}
 H = -\frac{\hbar^2}{2 m_e} \Bigl[ \nabla^2_1 + \nabla^2_2 + \nabla^2_3 \Bigr] - \frac{e^2}{r_{31}} - \frac{e^2}{r_{32}} - \frac{e^2}{r_{21}} \; \; \; \label{Hamil}
\end{eqnarray}
where $\hbar = \frac{h}{2 \pi}$ is the reduced Planck constant (also called Dirac constant) and $m_e$ is the electron mass and $- e$ is the electric charge of an electron. In this 
equation and everywhere below in this study the subscripts 1 and 2 designate two electrons ($e^{-}$), while the subscript 3 always denotes the positron ($e^{+}$) with the mass 
$m_{e}$ (the same electron mass) and positive electric charge $+e$, or $e$. In addition to the `numerical' indexes (1, 2, 3) in some cases we shall designate electrons by using the 
notaiton `-', while the notation `+' always means positively charged positron. In Eq.(\ref{Hamil}) the notations $r_{ij} = \mid {\bf r}_i - {\bf r}_j \mid = r_{ji}$ stand for three 
interparticle distances (= relative coordinates) which are the absolute values of differences of the Cartesian coordinates ${\bf r}_i$ of the three particles. Note that each relative 
coordinate $r_{ij}$ is a scalar which is rotationally and translationally invariant. However, these coordinates are not truly independent, since e.g., $\mid r_{32} - r_{31} \mid \le 
r_{21} \le  r_{32} + r_{31}$. This produces a number of problems for computations of three-particle integrals in these coordinates. To simplify such calculations it is better to apply
a set of three perimetric coordinates $u_1, u_2, u_3$ which are simply related to the relative coordinates: $u_{i} = \frac12 (r_{ik} + r_{jk} - r_{ij})$, while inverse relations take
the form $r_{ij} = u_i + u_j$. Three perimetric coordinates $u_1, u_2, u_3$ are independent of each other and each of them varies between 0 and $+ \infty$. The Jacobian of the transition 
$r_{jk} \rightarrow u_{i}$: $D_{u_1, u_2, u_3}(r_{32}, r_{31}, r_{21})$ is a constant which equals 2. 
    
Note also that in this study only atomic units $\hbar = 1, \mid e \mid = 1, m_e = 1$ are employed. In these units the explicit form of the Hamiltonian $H$, Eq.(\ref{Hamil}), is simplified
to the form
\begin{eqnarray}
 H = -\frac{1}{2} \Bigl[ \nabla^2_1 + \nabla^2_2 + \nabla^2_3 \Bigr] - \frac{1}{r_{31}} - \frac{1}{r_{32}} + \frac{1}{r_{21}} \; \; \; \label{Hamil1}
\end{eqnarray}
Note that the Hamiltonian, Eq.(\ref{Hamil1}), does not contain any ratio of masses and/or electric charges. It follows from here that the Ps$^{-}$ ion plays a central role in the general 
theory of Coulomb three-body systems with unit charges (for more details, see, e.g., \cite{Fro92}).        

To solve the non-relativistic Schr\"{o}dinger equation $H \Psi = E \Psi$ for the Ps$^{-}$ ion, where $E < 0$, and obtain highly accurate wave function(s) we approximate the unknown exact 
solution of the non-relativistic Schr\"{o}dinger equation with some efficient and fast convergent variational expansions. The best of such expansions is the exponential variational 
expansion in the relative coordinates $r_{32}, r_{31}, r_{21}$, or perimetric coordinates $u_1, u_2, u_3$. For the ground state of the Ps$^{-}$ ion the explicit form of this expansion is
\begin{eqnarray}
  \Psi &=& \frac12 ( 1 + \hat{P}_{12} ) \sum^{N}_{i=1} C_i \exp(-\alpha_i r_{32} - \beta_i r_{31} - \gamma_i r_{21}) \nonumber \\
  &=& \frac12 ( 1 + \hat{P}_{12} ) \sum^{N}_{i=1} C_i \exp[-(\alpha_{i} + \beta_{i}) u_{3} - (\alpha_{i} + \gamma_{i}) u_{2} - (\beta_{i} + \gamma_{i}) u_{3}) \; \; \; \label{exp}
\end{eqnarray}
where the notation $\hat{P}_{12}$ stands for the permutation operator of identical particles, $C_i$ ($i = 1, 2, \ldots, N$) are the linear parameters of the exponential expansion, 
Eq.(\ref{exp}), while $\alpha_i, \beta_i$ and $\gamma_i$ are the non-linear parameters of this expansion. The non-linear parameters must be varied in calculations to increase the overall 
efficiency and accuracy of the method. The best-to-date optimization strategy for these non-linear parameters was described in \cite{Fro2001}, while its modified version is presented in 
\cite{Fro2006}. The $3 N-$conditions $\alpha_{i} + \beta_{i} > 0, \alpha_{i} + \gamma_{i} > 0, \beta_{i} + \gamma_{i} > 0$ for $i = 1, 2, \ldots, N$ must be obeyed to guarantee 
covergence of all three-particle integrlas needed in computations. 

\section{Expectation values}

By using the highly accurate, variational wave function $\Psi$ constructed for the ground $1^1S-$state of the Ps$^{-}$ ion we can determine the expectation value of an arbitrary, in principle, 
self-adjoint operator $\hat{X}$. This is written in the following general form
\begin{equation}
  \langle \hat{X} \rangle = \frac{\langle \Psi \mid \hat{X} \mid \Psi \rangle}{\langle \Psi \mid \Psi \rangle} \label{expect}
\end{equation}
Formally, without loss of generality below we shall assume that our wave function has a unit norm, i.e. $\langle \Psi \mid \Psi \rangle = 1$ (see, discussion in \cite{Eps}). The total energy 
is the expectation value of the Hamiltonian $H$, Eq.(\ref{Hamil}), i.e. $E = \langle \Psi \mid \ H \mid \Psi \rangle$. The total energies $E$ of the ground $1^1S-$state of the Ps$^{-}$ ion 
determined for different trial wave functions can be found in Table I. Other possible choices of operators $\hat{X}$ in Eq.() lead to the different bound state properties, or properties, for 
short. A number of bound state properties were determined in earlier computations of the Ps$^{-}$ ion (see, e.g., \cite{BD1}, \cite{Ho}, \cite{Fro05} and references therein). In this study we 
present a large number of bound state properties of the Ps$^{-}$ ion determined to very high numerical accuracy (`essentially exact'). They can be found in Table II (in atomic units). Physical 
meaning of many of these properties is clear from the notations used in Tables I and II. For instance, the notation $\langle r_{ij} \rangle$ stands for the expectation value of the linear 
distance between particles $i$ and $j$. Another notation $\langle \delta_{ij} \rangle = \langle \delta({\bf r}_i - {\bf r}_j) \rangle$ denotes the expectation value of the (Dirac) delta-function 
between particles $i$ and $j$, while $\langle \delta_{321} \rangle = \langle \delta({\bf r}_i - {\bf r}_j) \delta({\bf r}_i - {\bf r}_k) \rangle$ is the expectation value of the triple 
delta-function. In general, the expectation value of each delta-function is the probability to locate two (or three) particles inside of one small sphere with the radius $R \approx \Lambda_e = 
\alpha a_0$, where $\Lambda_e$ is the Compton wavelength of electron, $\alpha = \frac{e^2}{\hbar c} \approx \frac{1}{137}$ is the fine structure constant and $a_0$ is the Bohr radius (see below). 

For the Ps$^{-}$ ion the expectation value of the electron-positron delta-function $\langle \delta_{+-} \rangle$ determines a number of annihilation rates, including the two- and three-photon 
annihilation rates (see below). The expectation value of the triple delta-function $\langle \delta_{321} \rangle = \langle \delta_{+--} \rangle$ is important to predict the one-photon annihilation 
rate $\Gamma_{1 \gamma}$. It is clear that some reliable criteria are needed to check the overall quality of the computed expectation values of delta-functions. In reality, we can introduce such 
criteria by considering the coincidence of the computed and predicted cusp values between each pair of particles. It was shown in early papers on Coulomb systems, including atoms and molecules 
\cite{Kato}, \cite{Schra}, that the following expectation value: 
\begin{equation}
 \nu_{ij} = \langle \hat{\nu}_{ij} \rangle = \frac{\langle \Psi \mid \delta({\bf r}_{ij}) \frac{\partial}{\partial r_{ij}} \mid \Psi \rangle}{\langle \Psi \mid \delta({\bf r}_{ij}) 
 \mid \Psi \rangle} \label{cusp}
\end{equation}
is always finite and its numerical value equals to $q_i q_j \frac{m_i m_j}{m_i + m_j}$, where $q_i, q_j$ are the corresponding electrical charges of particles, while $m_i, m_j$ are their masses. 
The expectation value $\nu_{ij}$ is called the cusp between two Coulomb particles $i$ and $j$. The coincidence of the computed expectation value of the cusp $\nu_{ij} = \langle \hat{\nu}_{ij} 
\rangle$ with its expected value, i.e. with $q_i q_j \frac{m_i m_j}{m_i + m_j}$, indicates the overall quality of the expectation value of the inter-particle delta-function. In actual applications 
to Coulomb systems this criterion works very well. Our computed and expected cusp values are presented in Table II. As one can see from Table II numerical coincidence between the predicted and 
computed expectation values of the electron-positron and electron-electron cusps can be considered as very good. The predicted value of the electron-positron cusp for the Ps$^{-}$ ion equals -0.5 
$a.u.$, while for the electron-electron cusp one finds 0.5 $a.u.$ Unfortunately, there is no similar criterion for the triple delta-function $\delta_{321}$, since the corresponding `three-particle 
cusp' is infinte for an arbitrary Coulomb system (see, e.g., \cite{Fock}, \cite{Fro99} and references therein). However, for trial functions with the finite number of regular (or non-singular)
basis functions the three-particle cusp $\nu_{123}$ can be defined (this value is finite) and used in computations \cite{Fro99}.   

The notations $\tau_{ij}$ stand for the expectation values of the interparticle cosine-functions which are defined traditionally:
\begin{equation}
 \tau_{ij} = \langle \cos({\bf r}_{ik}{}^{\wedge}{\bf r}_{jk}) \rangle = \langle \frac{{\bf r}_{ik}}{r_{ik}} \cdot \frac{{\bf r}_{jk}}{r_{jk}} \rangle
\end{equation}
where $(i, j, k) = (1, 2, 3)$, while the notation ${\bf r}_{ik}{}^{\wedge}{\bf r}_{jk}$ denotes the angle between the two vectors ${\bf r}_{ik}$ and ${\bf r}_{jk}$. For the Ps$^{-}$ ion there are 
two independent $\tau_{ij}$ values, i.e. $\tau_{21}$ (or $\tau_{--}$) and $\tau_{31}$ (= $\tau_{32}$, or $\tau_{+-}$). These two values are shown in Table II. For an arbitrary three-body system 
the sum of the $\tau_{21}, \tau_{31}$ and $\tau_{32}$ values is represented in the form
\begin{equation}
  \tau_{21} + \tau_{31} + \tau_{32} = 1 + 4 \langle f \rangle  \label{ff}
\end{equation}
where the notation $\langle f \rangle$ is the following expectation value  
\begin{eqnarray}
  \langle f \rangle &=& \int_0^{+\infty} \int_0^{+\infty} \int_{\mid r_{32} - r_{31} \mid}^{r_{32} + r_{31}} \Psi(r_{32}, r_{31}, r_{21}) 
  \frac{u_1 u_2 u_3}{r_{32} r_{31} r_{21}} \Psi(r_{32}, r_{31}, r_{21})  
  r_{32} r_{31} r_{21} dr_{32} dr_{31} dr_{21} \label{eq1} \\
 &=& 2 \int_0^{+\infty} \int_0^{+\infty}  \int_0^{+\infty} \Psi(u_{1}, u_{2}, u_{3}) u_1 u_2 u_3 \Psi(u_{1}, u_{2}, u_{3}) du_1 du_2 du_3 \nonumber 
\end{eqnarray}
where $r_{32}, r_{31}, r_{21}$ are the three relative coordinates and $u_1, u_2, u_3$ are the three perimetric coordinates, respectively. For the symmetric two-electron Ps$^{-}$ ion the equality, 
Eq.(\ref{ff}), takes the form $\tau_{21} + 2 \tau_{31} = 1 + 4 \langle f \rangle$. In applications to real systems the equality, Eq.(\ref{ff}), can be considered as an additional test of the trial 
wave function. Indeed, the expectation values in the both sides of Eq.(\ref{ff}) can be computed separately. The coincidence of the results in both sides of Eq.(\ref{ff}) indicates correctness of 
the procedure used. 

In general, there are a number of additional equalities for the expectation values determined for an arbitrary three-body system (or few-body system). For instance, consider the following identity
${\bf r}_{32} = {\bf r}_{31} + {\bf r}_{12}$. From this identity one finds $r^{2}_{32} = r^{2}_{31} + r^{2}_{21} - 2 {\bf r}_{31} \cdot {\bf r}_{21}$. This leads to the following identity for the 
expectation value $\langle {\bf r}_{31} \cdot {\bf r}_{21} \rangle$ 
\begin{equation}
    \langle {\bf r}_{31} \cdot {\bf r}_{21} \rangle = \frac12 (\langle r^{2}_{31} \rangle +  \langle r^{2}_{21} \rangle -  \langle r^{2}_{32} \rangle) \label{rrr}
\end{equation}
Two other analogous equalities for the $\langle {\bf r}_{32} \cdot {\bf r}_{12} \rangle$ and $\langle {\bf r}_{32} \cdot {\bf r}_{31} \rangle$ values can be derived by using the same approach. 
Another important example follows from the conservation of the momentum in an arbitrary three-body system. In the general case, for an arbitrary wave function $\Psi$ written in the relative 
coordinates we can write $({\bf p}_1 + {\bf p}_2 + {\bf p}_3) \Psi = 0$. It is assumed here that the wave function $\Psi$ depends upon three relative coordinates which are translationally (and 
rotationally) invariant. From this equation one finds three different identies between corresponding expectation values, e.g., 
\begin{equation}
    \langle {\bf p}_{1} \cdot {\bf p}_{2} \rangle = \frac12 (\langle p^{2}_{3} \rangle - \langle p^{2}_{1} \rangle - \langle p^{2}_{2} \rangle) \label{ppp}
\end{equation}
and
\begin{equation}
    \langle {\bf p}_{1} \cdot {\bf p}_{3} \rangle = \frac12 (\langle p^{2}_{2} \rangle - \langle p^{2}_{1} \rangle - \langle p^{2}_{3} \rangle) \label{ppp1}
\end{equation}
For the Ps$^{-}$ ion the first and second expectation values in the right-hand side of Eq.(\ref{ppp1}) are equal to each other (particles 1 and 2 are identical electrons). Therefore, in this case 
one finds $\langle {\bf p}_{1} \cdot {\bf p}_{3} \rangle = \langle {\bf p}_{2} \cdot {\bf p}_{3} \rangle = -\langle p^{2}_{3} \rangle < 0$. In atomic units for the Ps$^{-}$ ion we have ${\bf p}_k 
= -\imath \nabla_k$ ($k$ = 1, 2, 3), where $\imath$ is the imaginary unit and $\nabla_k$ is the gradient operator for $k-$th particle (also called the Hamilton operator). Additional relations 
between other expectation values can be investigated analogously.

Note that the current accuracy achieved in calculations of a large number of bound state properties of the Ps$^{-}$ ion is very high. This follows from Table II where only stable decimal digits 
are shown for each property. Briefly, we can say that many bound state properties of the Ps$^{-}$ ion are now known to the level which cannot easily be improved in future computations, or such an 
improvement has no direct physical sense. By using the expectation values from Table II we can evalauate some fundamental properties of the Ps$^{-}$ ion. The main interest for the Ps$^{-}$ ion is 
related to the positron annihilation, or annihilation of the $(e^{-}, e^{+})-$pair(s). In reality, we have a number of different few-photon annihilation rates, including two- and three-photon 
annihilation rates. These values are evaluated below. Note also that numerical evaluations of the relativistic and lowest-order QED corrections for the Ps$^{-}$ ion can be found in \cite{Fro05} 
(see also \cite{Fro07a}). It should be mentioned, however, that such evaluations were not completed. Moreover, some of the values determined in \cite{Fro05} must be re-calculated to better accuracy 
and with the use of different algorithms. It is also clear that by using our current methods we cannot finish our numerial evalution of the lowest-order QED correction for the Ps$^{-}$ ion, since 
this ion has no infinitely heavy, central nucleus. This problem is briefly discussed in the Conclusion.

\section{Positron annihilation}

Now, let us discuss the process of positron annihilation, or annihilation of the electron-positron pair(s), in the three-body Ps$^-$ ion. As mentioned above this process is of great interest in 
numerous applications, including astrophysics, solid state physics, etc. It is also important for workability of many technical devices, e.g., modern communication devices. In the general case, 
annihilation of the electron-positron pair(s) in the Ps$^{-}$ ion proceeds with the emission of a number of photons (see, e.g., \cite{AB}, \cite{Grein}), e.g.,
\begin{eqnarray}
 e^{+} + e^{-} = \gamma_1 + \gamma_2 + \ldots + \gamma_K \; \; \; \label{eq0} 
\end{eqnarray}
where $\gamma_k$ ($k = 1, \ldots, K$) are the emitted photons and $K$ is the maximal number of such photons. Each of the annihilation processes has its unique annihilation width, or annihilation 
rate $\Gamma_{k \gamma}$. For the Ps$^{-}$ ion the following annihilation rates are important in applications: $\Gamma_{2 \gamma}, \Gamma_{3 \gamma}, \Gamma_{4 \gamma}, \Gamma_{5 \gamma}$ and 
$\Gamma_{1 \gamma}$ (here they are ordered by their numerical values). Formulas for these annihilation rates were discussed in a number of earlier studies (see, e.g., \cite{Fro05} and 
\cite{Fro07a} and references therein). For instance, the known analytical expression for the $\Gamma_{2 \gamma}$ rate is written in the form 
\begin{eqnarray}
 \Gamma_{2 \gamma}({\rm Ps}^{-}) = n \pi \alpha^4 c a^{-1}_0 \Bigl[ 1 - \frac{\alpha}{\pi} \Bigl( 5 - \frac{\pi^2}{4} \Bigr)\Bigr] \langle \delta({\bf r}_{+-}) \rangle \approx 100.34560545419 
 \cdot 10^{9} \langle \delta_{+-} \rangle \; sec^{-1} \label{An2g}
\end{eqnarray}
where $\alpha = \frac{e^2}{\hbar c} = 7.2973525698 \cdot 10^{-3} \Bigl(\approx \frac{1}{137}\Bigr)$ is the dimensionless fine structure constant, $c = 2.99792458 \cdot 10^{8}$ $m \cdot sec^{-1}$ 
is the speed of light in vacuum, and the Bohr radius $a_0$ equals $0.52917721092 \cdot 10^{-10}$ $m$ \cite{CRC}. Also, in this formula $\hbar = \frac{h}{2 \pi}$ is the reduced Planck constant (or
Dirac constant) and $n$ is the total number of electron-positron pairs in the polyelectron $e^{+}_p e^{-}_q$, i.e. $n = p q$. For the three-body Ps$^{-}$ ion we have $n = 2$. In Eq.(\ref{An2g}) 
and in formulas below the notation $\langle \delta_{+-} \rangle$ stands for the expectation value of the electron-positron delta-function determined (in atomic units) for the $1^1S$-state of the 
Ps$^-$ ion. The formula, Eq.(\ref{An2g}), also includes the lowest-order radiative correction \cite{Brown} (or QED correction in modern language). By using this formula for the $\Gamma_{2 \gamma}$ 
rate we can evaluate the four-photon annihilation rate $\Gamma_{4 \gamma}$, since the relation between these two values takes the form \cite{Fro09} 
\begin{equation}
 \Gamma_{4 \gamma}({\rm Ps}^-) \approx 0.274 \Bigl(\frac{\alpha}{\pi}\Bigr)^2 \Gamma_{2 \gamma}({\rm Ps}^-) = 1.4783643 \cdot 10^{-6} \Gamma_{2 \gamma}({\rm Ps}^-)  \label{An4g}
\end{equation}
This formula is based on the result from \cite{PRA83} for an isolated electron-positron pair $(e^{+}, e^{-})$. The idea that all annihilation rates with even number of photons are related with
each other was proposed by Ferrante in 1969 \cite{Fer}. Later, it was found that this idea is correct in application to the Ps$^{-}$ ion and other polyelectrons. Furthermore, the same idea is 
also correct for annihilation rates with odd numbers of the emitted photons.    

The three-photon annihilation rate $\Gamma_{3 \gamma}({\rm Ps}^{-})$ is $\approx$ 1000 times smaller than the two-photon annihilation rate $\Gamma_{2 \gamma}({\rm Ps}^{-})$. The corresponding
analytical expression for $\Gamma_{3 \gamma}({\rm Ps}^-)$ is \cite{Fro05}
\begin{eqnarray}
 \Gamma_{3 \gamma}({\rm Ps}^-) = n \frac{4 (\pi^2 - 9)}{3} \alpha^5 c a^{-1}_0 \langle \delta({\bf r}_{+-}) \rangle \approx 2.7185459576 \cdot 10^8 \langle \delta_{+-} \rangle \; 
  sec^{-1} \; \; \; , \label{An3g}
\end{eqnarray}
This formula allows one to evaluate the five-photon annihilation rate in the Ps$^-$ ion. Indeed, by using the formula from \cite{PRA83} one finds the following result
\begin{equation}
 \Gamma_{5 \gamma}({\rm Ps}^-) \approx 0.177 \Bigl(\frac{\alpha}{\pi}\Bigr)^2 \Gamma_{3 \gamma}({\rm Ps}^-) = 0.955001778 \cdot 10^{-6} \Gamma_{3 \gamma}({\rm Ps}^-)   \label{An5g}
\end{equation}
As follows from Eq.(\ref{An4g}) and Eq.(\ref{An5g}) the $\Gamma_{4 \gamma}({\rm Ps}^-)$ and $\Gamma_{5 \gamma}({\rm Ps}^-)$ annihilation rates are substantially (in $\approx 10^5$ times) less
than the corresponding $\Gamma_{2 \gamma}({\rm Ps}^-)$ and $\Gamma_{3 \gamma}({\rm Ps}^-)$ annihilation rates. 

In the three-particle Ps$^{-}$ ion one can also observe the process of one-photon annihilation of the electron-positron pair. For an isolated electron-positron pair the one-photon annihilation is strictly 
prohibited. However, a close presence of a third particle drastically changes this situation. The one-photon annihilation rate (or width) $\Gamma_{1 \gamma}$ is written in the form \cite{Kru} (see also 
\cite{Fro05} and \cite{Chu})
\begin{eqnarray}
 \Gamma_{1 \gamma} = \frac{64 \pi^2}{27} \cdot \alpha^8 \cdot c \cdot a_0^{-1} \cdot \langle \delta_{321} \rangle = 1065.756921658 \cdot \langle \delta_{321} \rangle  \; \; \; sec^{-1} 
 \; \; \; , \nonumber
\end{eqnarray}
where the notation $\langle \delta_{321} \rangle$ stands the expectation value of the triple delta-function (in atomic units) computed for the ground state of the Ps$^-$ ion. This value is the probability 
of finding all three-particles at one spatial point.   

The sum of all partial annihilation rates, i.e. $\Gamma_{2 \gamma}({\rm Ps}^{-}), \Gamma_{3 \gamma}({\rm Ps}^{-}), \Gamma_{4 \gamma}({\rm Ps}^{-}), \ldots$ is the total annihilation rate, or
$\Gamma({\rm Ps}^{-})$. For the Ps$^{-}$ ion we have
\begin{eqnarray}
 \Gamma &=& \Gamma_{2 \gamma}({\rm Ps}^{-}) + \Gamma_{3 \gamma}({\rm Ps}^{-}) + \Gamma_{4 \gamma}({\rm Ps}^{-}) + \ldots \approx \Gamma_{2 \gamma}({\rm Ps}^{-}) + \Gamma_{3 \gamma}({\rm Ps}^{-}) 
  \nonumber \\
 &=& 2 \pi \alpha^4 c a^{-1}_0 \Bigl[ 1 - \alpha \Bigl( \frac{17}{\pi} - \frac{19 \pi}{12} \Bigr)\Bigr] \langle \delta({\bf r}_{+-}) \rangle \approx 100.617460050 
 \cdot 10^{9} \langle \delta_{+-} \rangle \; sec^{-1} \label{Annht}
\end{eqnarray}   
As follows from these formulas to determine all partial and total annihilation rates we need to know the expectation values of the electron-positron delta-function $\langle \delta_{+-} \rangle$ and 
three-particle delta-function $\langle \delta_{321} \rangle$ for the ground $1^1S-$state of the Ps$^{-}$ ion. By using the expectation values of these delta-functions from Table II we have found the 
following numerical values of all mentioned annihilation rates: $\Gamma_{2 \gamma} = 2.08048530684 \cdot 10^9$, $\Gamma_{3 \gamma} = 5.6364151625 \cdot 10^6$, $\Gamma_{4 \gamma} = 3.0757152758 \cdot 
10^3$, $\Gamma_{5 \gamma} = 5.382786501$, $\Gamma_{1 \gamma} = 3.82491558 \cdot 10^{-2}$ and $\Gamma = 2.08612172200 \cdot 10^9$ (all values are given in $sec^{-1}$). These values are slightly better than 
our previous values given in \cite{Fro09}. Different aspects of the positron annihilation of the the Ps$^{-}$ ion were discussed in a large number of earlier studies (see, e.g., \cite{BD1}, \cite{Ho}, 
\cite{Chu}, \cite{Kru}, \cite{Czam} and others).
 
\section{Photodetachment}

Photodetachment of the Ps$^{-}$ ion is of great interest in applications to astrophysics and propagation of radiation in our Galaxy. As is well known the center of our Galaxy contains a number of sources 
of the annihilation $\gamma-$quanta with $E_{\gamma} \approx$ 0.511 $MeV$ (see, e.g., \cite{Dra} and references therein). This indicates the presence of objects with very high (local) temperatures $T \ge$ 
350 - 400 $keV$ and formation of large numbers of the electron-positron pairs $(e^{-},e^{+})$, Ps$^{-}$ and Ps$^{+}$ ions, bi-positronium Ps$_2$ and other polyelectron species. Photodetachment of the 
Ps$^{-}$  ion and other polyelectrons leads to very intense absorbtion of the infrared radiation in such spatial areas. Photodetachment of the Ps$^{-}$ ion(s) was considered in \cite{BD2}, \cite{Shim}. 
Some closely related problems, e.g., elastic electron-positronium scattering and photodetachment of the Ps$^{-}$ ion by a model Yukawa-type potential, were discussed in \cite{Hum} and \cite{Ho08}, 
respectively. There is a well known experimenta paper about photodetachment of the Ps$^{-}$ ion \cite{Tachi}. 

To simplify theoretical analysis and numerical calculations in this study we shall apply an effective method of `asymptotic photodetachment', or photodetachment of the Ps$^{-}$ ion at very large distances 
form the `geometrical center' of this three-body system. This method was originally proposed by Hans Bethe in 1935 when he considered photodetachment of the deuterium nucleus. This method is based on the 
fact that the wavelength of the `acting light' $\lambda$ is much larger than the effective geometrical size of the system $a$, i.e. we always have $\lambda \gg a$. Therefore, we can use the wave function of 
the Ps$^{-}$ ion in its asymptotic form which is defined only at very large (or asymptotic) distances between outer-most electron $e^{-}$ and neutral central cluster Ps. Analytical form of the wave function
of the Ps$^{-}$ ion can be found from the following formula
\begin{equation}
 \mid \Psi(r) \mid = C r^{\frac{Z}{\gamma} - 1} \exp(-\gamma r) \label{ph1}
\end{equation}
where $C$ is some numerical constant, $Z = Q - N_e + 1$, where $Q$ is the electric charge of the nucleus and $N_e$ is the total number of bound electrons. The parameter $\gamma$ in this equation equals 
$\gamma = \sqrt{2 I_1}$ and $I_1 = \chi_1$ is the (first) ionization potential which corresponds to the dissociation of the Ps$^{-}$ ion, i.e. Ps$^{-}$ = Ps + $e^{-}$. For the Ps$^{-}$ ion one finds in 
Eq.(\ref{ph1}): $Q = 1, N_e = 2$, and therefore, $Z = 0$ and the long range asymptotic of the wave function, Eq.(\ref{ph1}), is represented in the Yukawa-type form $\mid \Psi(r) \mid = \frac{C}{r} \cdot
\exp(-\gamma r)$, where $C$ is a constant which must provide the best correspondence of Eq.(\ref{ph1}) with the highly accurate wave function of the Ps$^{-}$ ion at large $r$. The highly accurate wave 
function of the Ps$^{-}$ ion is assumed to be known from numerical computations of this ion. 

The photodetachment cross-section $\sigma($Ps$^-)$ of the Ps$^-$ ion is written in the following form (derivation of this formula is discussed in \cite{Fro2014})    
\begin{eqnarray}
 d\sigma({\rm H}^-) = \alpha a^{2}_{0} \frac{p_e}{3 \pi \omega} \mid {\bf e}_f \cdot \int \psi^{*}_f \Bigl(\frac{\partial}{\partial {\bf r}} \psi_{i}\Bigr) d^3{\bf r} \mid^2 do
 = \alpha a^{2}_{0} \frac{p_e}{3 \pi \omega} \mid {\bf e}_f \cdot \int \psi_i \Bigl(\frac{\partial}{\partial {\bf r}} \psi^{*}_{f}\Bigr) d^3{\bf r} \mid^2 do \; \; \; \label{cr-sec}
\end{eqnarray}
where $do = \sin \theta d\theta d\phi$ is an elementary volume in spherical coordinates. The normalized wave function of the incident state is $C r^{\frac{Z}{t} - 1} \cdot \exp(-\gamma r)$, while the wave 
function of the final state is $\exp( \imath {\bf p}_e \cdot {\bf r})$. By substituting these exressions into Eq.(\ref{cr-sec}) one finds
\begin{eqnarray}
 d\sigma({\rm Ps}^-) &=& \alpha a^{2}_{0} \Bigl( \frac{128 \pi p^{3}_e}{3 \omega} \Bigr) ({\bf e}_f \cdot {\bf n}_e)^2 \cdot \mid \frac{1}{p_e} \int_{0}^{+\infty} \psi_i(r) \sin(p_e r) 
 r dr \mid^2 do \nonumber \\
 &=& \alpha a^{2}_{0} \frac{128 \pi p^{3}_e}{3 \omega} ({\bf e}_f \cdot {\bf n}_e)^2 \cdot \mid \frac{1}{p_e} \int_{0}^{+\infty} \exp(-\gamma r) \sin(p_e r) dr \mid^2 do
 \; \; \; \label{cr-sec1}
\end{eqnarray}
where ${\bf p}_e = p_e {\bf n}_e$ and ${\bf n}_e$ is the unit vector which determines the direction of propagation of the emitted photo-electron. Here and below $\alpha$ is the fine structure constant, 
$a_0$ is the Bohr radius and $\gamma = \sqrt{2 I_1}$. The (first) ionization potential $I_1$ of the Ps$^{-}$ ion is proportional to the difference of the total ground state energies of the Ps$^{-}$ ion 
(see Table I) and Ps two-body system (-0.25 $a.u.$ exactly). The exact expression also contains the factor $\frac23$, i.e. $I_1 = \frac23 (0.262005070\ldots - 0.25)$ and $\gamma =  \sqrt{\frac43 
(0.262005070\ldots - 0.25)}$. A few steps of additional transformations (see, e.g., \cite{Fro2014}) lead to the following final formula (in $a.u.$ $\omega = \frac32 p^2_e + \frac32 \gamma^2$)
\begin{eqnarray}
 d\sigma({\rm Ps}^-) = \frac{128 \pi}{9} \cdot C^2 \alpha a^{2}_{0} \Bigl( \frac{p_e}{p^{2}_e + \gamma^2} \Bigr)^3 ({\bf k}_f \times {\bf n}_e)^2 do = 
 \frac{256 \pi^2}{9} \cdot C^2 \alpha a^{2}_{0} \Bigl( \frac{p_e}{p^{2}_e + \gamma^2} \Bigr)^3 \sin^2\Theta d\Theta \; \; \; \label{cr-sec2}
\end{eqnarray}
where $\sin\Theta = {\bf k}_f \times {\bf n}_e$ and both vectors ${\bf k}_f$ and ${\bf n}_e$ have unit norm. For the total cross-section one finds
\begin{eqnarray}
 \sigma({\rm Ps}^-) = \frac{512 \pi^2}{9} \cdot C^2 \alpha a^{2}_{0} \Bigl( \frac{p_e}{p^{2}_e + \gamma^2} \Bigr)^3 \approx 38.2449007 \cdot 10^{-18} \cdot C^2 \Bigl( \frac{p_e}{p^{2}_e + \gamma^2} \Bigr)^3
 \; \; \; \label{cr-sec3}
\end{eqnarray}
where $C$ is expressed in atomic units. The formula which allows one to determine the constant $C$ takes the from
\begin{eqnarray}
  C = D^{-\frac12}_a R \exp(\gamma R) \Psi(R, 0, R) &=& D^{-\frac12}_a R \exp(\gamma R) D^{-\frac12}_r \sum^{N}_{i=1} C_i \Bigl[\exp(-\alpha_i R - \gamma_i R) \; \; \; \label{const} \\ 
  &+& \exp(-\beta_i R - \gamma_i R)\Bigr] \nonumber
\end{eqnarray}
where $D_r$ is the normalization constant for the radial part of the total wave function of the ground state of the Ps$^{-}$ ion, while $D_a$ is the normalization constant for the angular part of this
wave function. However, there is an additional normalization constant which is defined for the angular part of the total wave function. This constant equals $D_a = \frac{1}{\sqrt{8 \pi}}$. In our 
calculations with the use of 700 basis functions, Eq.(\ref{exp}), we have found that the constant $C$ equals $C \approx 0.9322567 \cdot \frac{1}{\sqrt{8 \pi}}$. Now from Eq.(\ref{const}) one finds $C$ 
= 0.18595831 (compare with \cite{BD2}) and expression for the photodetachment cross-section takes the form 
\begin{eqnarray}
 \sigma({\rm Ps}^-) = A \cdot \Bigl( \frac{p_e}{p^{2}_e + \gamma^2} \Bigr)^3 = 1.3225275 \cdot 10^{-18} \cdot \Bigl( \frac{p_e}{p^{2}_e + \gamma^2} \Bigr)^3 \; \; \; cm^{2} \; \; \; \label{cr-sec4}
\end{eqnarray}
Analogous numerical computations with the use of 3500 basis functions (see Table III) lead to the following value of $C \approx 0.1859599866$ (at $R = 70.5 a.u.$). The photodetachment cross-section is 
written in the form 
\begin{eqnarray}
 \sigma({\rm Ps}^-) = 1.322555 \cdot 10^{-18} \cdot \Bigl( \frac{p_e}{p^{2}_e + \gamma^2} \Bigr)^3 \; \; \; cm^{2} \; \; \; \label{cr-sec5}
\end{eqnarray}

The formulas Eqs.(\ref{cr-sec4}) - (\ref{cr-sec5}) solve the problem of the photodetachment of the Ps$^{-}$ ion in the `asymptotic approximation', which is sufficient for various applications in 
astrophysics. Note that the general approach to the photodetachment of the Ps$^{-}$ ion, i.e. approach which is not based on the `asymptotic', long-range approximation of the wave function, has not been 
developed yet. In real applications this `asymtotic' approach can be generalized to include more complex cases of the photodetachment of the Ps$^{-}$ ion, including semi-relativistic case and 
photodetachment when in the final state one finds a free electron and secondary photon of smaller energy.
 
\section{Conclusion}

The bound state properties of the ground state in the Ps$^{-}$ ion are investigated with the use of highly accurate results from recent calculations. At this moment we can conclude that the positron 
annihilation in the Ps- ion is a well studied phenomenon. In the next studies we plan to consider some unsolved problems which are currently known for the Ps$^{-}$ ion. A special attention will be 
given to calculation of the lowest-order QED corrections for the ground (bound) state of the Ps$^{-}$ ion. The lowest-order relativistic and QED corrections for the Ps$^{-}$ were considered in 
\cite{Fro05}, \cite{BD5} and \cite{Drake05}. In particular, in our earlier study \cite{Fro05} we have determined the Bethe logarithm for the ground state in the Ps$^{-}$ ion ($ln K_0 ($Ps$^{-}$) = 
3.00502533(5)). The two (singular) expectation values $\langle r^{-3}_{+-} \rangle$ and $\langle r^{-3}_{--} \rangle$ for the Ps$^{-}$ ion are also known to very good numerical accuracy (see, e.g., 
\cite{Fro07a} and/or Table II). However, by using these values and Bethe logarithm we cannot determine the lowest-order QED correction for the Ps$^{-}$ ion in the same way as we did for tow-electron atoms 
and ions (see, e.g., \cite{JCP2014}). The reason is obvious, since the corresponding `small parameter', i.e. the ratio of the electron and positronium masses ($m_e$ and $M_{{\rm Ps}}$, where Ps is the 
central cluster), equals $\frac12$. The corresponding series for the lowest-order QED correction converges very slow. Currently, it is clear that any actual progress in computation of the lowest-order QED 
correction for the ground state in the Ps$^{-}$ ion can be based on the derivation of the Bethe –- Salpeter equation for weakly-bound non-Coulomb three-body systems \cite{BS} (see also discussion in 
\cite{Grein}).

\newpage
 \begin{table}[tbp]
   \caption{The total energies $E$ of the Ps$^{-}$ ion (in atomic units).}
     \begin{center}
     \begin{tabular}{| c | c | c |}
      \hline\hline
  $K$ & $E($Ps$^{-}$; variant A) & $E($Ps$^{-}$; variant B) \\
     \hline\hline
 3500 & -0.26200507 02329801 07770398 027 & -0.26200507 02329801 07770398 256 \\

 3700 & -0.26200507 02329801 07770399 455 & -0.26200507 02329801 07770399 614 \\

 3800 & -0.26200507 02329801 07770400 032 & -0.26200507 02329801 07770400 078 \\

 3840 & -0.26200507 02329801 07770400 250 & -0.26200507 02329801 07770400 279 \\
        \hline\hline
  \end{tabular}
  \end{center}
  \end{table}
%


\begin{table}[tbp]
   \caption{The expectation values of some propeties (in atomic units) for the Ps$^{-}$ ion. The notations `+' and `3' designate positron, while
            the notations `-' and `1' and `2' stand for electrons.}
     \begin{center}
     \begin{tabular}{| c | c | c | c |}
      \hline\hline
 $\langle r^{-2}_{+-} \rangle$ & $\langle r^{-2}_{--} \rangle$ & $\langle r^{-1}_{+-} \rangle$ & $\langle r^{-1}_{--} \rangle$ \\
      \hline 
 0.279326542224953365 & 0.0360220584545367961 & 0.33982102305922030648 & 0.15563190565248039742 \\
       \hline
 $\langle r_{+-} \rangle$ & $\langle r_{--} \rangle$ & $\langle r^{2}_{+-} \rangle$ & $\langle r^{2}_{--} \rangle$ \\
      \hline
 5.48963325235944993332 & 8.54858065509918611146 & 48.4189372262379554122 & 93.178633847981329005 \\
      \hline
 $\langle r^{3}_{+-} \rangle$ & $\langle r^{3}_{--} \rangle$ & $\langle r^{4}_{+-} \rangle$ & $\langle r^{4}_{--} \rangle$ \\
      \hline
 607.295629623278442191 & 1265.5804478781441204 & 9930.638679796004154 & 21054.45338925835809 \\
     \hline \hline
 $\langle [r_{32} r_{31}]^{-1} \rangle$ & $\langle [r_{+-} r_{--}]^{-1} \rangle$ & $\langle [r_{32} r_{31} r_{21}]^{-1} \rangle$ & $\langle \delta({\bf r}_{+--}) \rangle$ \\
     \hline
 0.090935346529989403768 & 0.060697690288581955165 & 0.0220342380163358569 & 3.58891929212$\cdot 10^{-5}$ \\
     \hline\hline
 $\langle \delta({\bf r}_{+-}) \rangle$ & $\nu_{+-}^{(a)}$ & $\langle \delta({\bf r}_{--}) \rangle$ & $\nu_{--}^{(a)}$ \\
      \hline
 0.020733198005219 & -0.500000000057701 & 1.7099675635500$\cdot 10^{-4}$ & 0.4999999981459 \\
     \hline
 $\tau_{eN}$ & $\tau_{ee}$ & $\langle f \rangle$ & $\langle {\bf r}_{31} \cdot {\bf r}_{32} / r^3_{31} \rangle$ \\
      \hline
 0.5919817011489022333 & 0.01976963281713200176 & 0.0509332587787341171 & -0.1234320911052318 \\
     \hline\hline 
 $\langle -\frac12 \nabla^2_{-} \rangle$ & $\langle -\frac12 \nabla^2_{+} \rangle$ & $\langle \nabla_1 \cdot \nabla_2 \rangle$ & $\langle \nabla_{+} \cdot \nabla_{-} \rangle$ \\
      \hline
 0.0666192945358900085 & 0.1287664811612000907 & -4.47210791057992633$\cdot 10^{-3}$ & -0.1287664811612000907 \\
     \hline
 $\langle {\bf r}_{+-} \cdot {\bf r}_{--} \rangle$ & $\langle (r^{-3}_{+-})_R \rangle$ & $\langle (r^{-3}_{--})_R \rangle$ & $\langle r^{-3}_{+-} \rangle$ \\
      \hline
 46.5893169239906645026 & -0.25348417470871 & 0.01131050073132 & 7.056875445764$\cdot 10^{-3}$ \\                                          
    \hline \hline
  \end{tabular}
  \end{center}
 ${}^{(a)}$The expected cusp values (in $a.u.$) for the Ps$^{-}$ ion are $\nu_{+-} = -0.5$ and $\nu_{--} = 0.5$ (exactly).
  \end{table}
%

\begin{table}[tbp]
   \caption{Numerical constant $C$ from Eqs.(22) (23) determined for different electron-positronium distances $R$ (in atomic units) for the ground state of the Ps$^{-}$ ion.
            The constant $A$ is defined in Eq.(24).}
     \begin{center}
     \begin{tabular}{| c | c | c | c |}
      \hline\hline
  $R$ & $C$ &  $\frac{C}{\sqrt{8 \pi}}$ & $A$ \\
     \hline\hline
  65.5 & 0.93224641707 & 0.1859562558 & 1.322502E-18 \\

  66.0 & 0.93224854366 & 0.1859566800 & 1.322508E-18 \\

  66.5 & 0.93225060763 & 0.1859570917 & 1.322514E-18 \\

  67.0 & 0.93225261127 & 0.1859574913 & 1.322519E-18 \\

  67.5 & 0.93225455679 & 0.1859578794 & 1.322525E-18 \\

  68.0 & 0.93225644627 & 0.1859582563 & 1.322530E-18 \\

  68.5 & 0.93225828175 & 0.1859586224 & 1.322535E-18 \\

  69.0 & 0.93226006515 & 0.1859589782 & 1.322540E-18 \\

  69.5 & 0.93226179830 & 0.1859593239 & 1.322545E-18 \\

  70.0 & 0.93226348299 & 0.1859596599 & 1.322550E-18 \\

  70.5 & 0.93226512090 & 0.1859599866 & 1.322555E-18 \\

  71.0 & 0.93226671366 & 0.1859603043 & 1.322559E-18 \\
        \hline\hline
  \end{tabular}
  \end{center}
  \end{table}
\end{document}